\documentclass[11pt,twocolumn]{article}
\usepackage[utf8]{inputenc}

\usepackage{authblk}

\usepackage{tikz}
\usetikzlibrary{decorations.pathreplacing,calc}
\usepackage{colortbl}
\usepackage{pgfplots,pgfplotstable}
\definecolor{cellshade}{rgb}{0.42, 0.55, 0.84}
\newcolumntype{C}[1]{>{\centering\arraybackslash}m{#1}}

\usepackage{amssymb}

\usepackage{xcolor}
\usepackage{listings}
\usepackage{url}

\begin{document}

\title{Simple Yet Efficient Content Based Video Copy Detection}

\author[1]{J\"org P. Bachmann}
\author[2]{Benjamin Hauskeller}
\affil[1]{ \texttt{joerg.bachmann@informatik.hu-berlin.de}}
\affil[2]{ \texttt{hauskelb@informatik.hu-berlin.de}}
\affil[1,2]{ Humboldt-Universität zu Berlin, Germany}

\maketitle

\begin{abstract}

Given a collection of videos, how to detect content-based copies efficiently with high accuracy?
Detecting copies in large video collections still remains one of the major challenges of multimedia retrieval.
While many video copy detection approaches show high computation times and insufficient quality, we propose a new efficient content-based video copy detection algorithm improving both aspects.
The idea of our approach consists in utilizing self-similarity matrices as video descriptors in order to capture different visual properties.
We benchmark our algorithm on the MuscleVCD ST1 benchmark dataset and show that our approach is able to achieve a score of 100\% and a score of at least 93\% in a wide range of parameters.

\end{abstract}

\section{Introduction}

Nowadays, a vast amount of video data is uploaded and shared on community sites such as YouTube or Facebook.
This leads to many problem statements such as copyright protection, duplicate detection, analyzing statistics of particular broadcast advertisements, or simply searching for large videos containing certain scenes or clips.
Two basic approaches exist to address these challenges, namely watermarking and content-based copy detection (CBCD).
Watermarking suffers from being vulnerable to transformations frequently performed during copy creation of a video (e.g.\ resizing or reencoding).
Furthermore, watermarking cannot be used on videos unmarked before distribution.
In contrast, CBCD is about finding copies of an original video by specifically comparing the contents and is thus more robust against transformations done during copy creation.
These transformations include resolution and encoding changes, cropping, blurring, and insertion of logos.
Hence, copies are near-duplicates and it is natural to use a distance or similarity function to discover them.

Of course, the distance function needs to be robust against the transformations mentioned above.
On the other hand, it has to discriminate videos from different sources to reduce the false alarm rate.
In this paper, a video descriptor is proposed which leads to a distance function, successfully addressing these goals.

\subsection{Related Work}

Although we cannot give a full overview on this topic, we like to roughly classify the existing approaches related to CBCD.
Videos are sequences of images or frames, hence comparing videos is based on comparing images.
Many approaches compare global features created per frame \cite{fastsequencematching, mpeg7, vstringeditdist}.
These global features include mean color values and color histograms.
To achieve higher discriminability, each image is partitioned into a grid and global features are calculated for each block of the grid \cite{Yeh2009, Chen2008, fastsequencematching}.
In contrast to global features, local features are more robust against transformations when searching for similar images \cite{efficientnear-duplicate}.
For example, Harris Corner Detectors are used to create feature descriptors in \cite{zgridcbcd} and \cite{scalablevideodbmining}.
Since these global and local feature descriptors are created on a per frame basis, they are called spatial descriptors.

For video CBCD, temporal information also needs to be taken into account (see \cite{Chen2008, spatiotemporaldementhon, Yeh2009, fastsequencematching, mpeg7, vstringeditdist}), leading to so-called spatio-temporal descriptors.
One approach compares temporal ordinal rankings of global features \cite{Chen2008, signaturebyordmeasure, Harvey2012}.
Since ordinal rankings depend on a total order of the features, they cannot be built for more robust local features.
A new concept called \textit{video strands} indicating movement of colored regions within half-second segments was introduced as space-time descriptors by the authors in \cite{spatiotemporaldementhon}.
The edit distance was first used to compare videos in 1998 \cite{raey, vstringeditdist}, where the global features of frames are quantized to obtain strings over an alphabet, thus the edit distance can be applied to compare two videos.
Recently, different derivations of this technique have been developed \cite{Yeh2009, fastsequencematching, mpeg7}.

The here proposed video descriptor is a derivative of the self-similarity matrix which is a common tool for structural analysis of time series.
The main purpose of the self-similarity matrix is to create recurrence plots.
Thus, similar subsequences of one time series can be found \cite{recurrenceplots, recurrenceplotscomplex}.
An example for an application is the structural analysis of music track arrangements \cite{transpositioninvariant, visualizingmusic}.
However, we do not use the self-similarity for analyzing a time series but we do use the (reduced) self-similarity matrices of two time series to assess their similarity.

\subsection{Contribution and Organization}

A contribution of our spatio-temporal video descriptor (see Section\ \ref{theory}) is its modularity, i.e.\ the separation of temporal and spatial aspects.
It uses an arbitrary distance or similarity function comparing two images to create the descriptor for a video.
Hence, the user is free to choose any suitable distance function out of a large set already defined by the research community \cite{cbirearlyyears} describing the distances between images properly.
Due to this flexibility, we were able to improve our simple yet fast image distance function to achieve a score of 100\% in the MuscleVCD ST1 benchmark.
Comparing two videos is then performed by comparing these descriptors without calling the distance function for any image again.
Thus, the execution time of comparing two videos at runtime does not depend on the complexity of the underlying image distance function, even if images are compared using expensive, robust, and discriminating image distance functions.
With this descriptor we achieve to create a distance function fulfilling all desired requirements described above.

The remainder of this paper is organized as follows.
The proposed video distance function as well as the CBCD decision algorithm is motivated and defined in Section~\ref{theory}.
Experiments evaluate its efficiency in Section~\ref{experiments}.
Section~\ref{conclusion} focuses on conclusions and future work.

\section{Theoretical Foundation}
\label{theory}

\subsection{Preliminaries}

To formalize the problem of CBCD, we denote the space of all (gray-scale) images with $\mathbb I$ and assume its pixel values in a range from $0$ to $1$.
A video is a sequence $V=(v_0,\ldots,v_{n-1})$ of images $v_i\in\mathbb I$ with equal resolution.
We denote $|V|=n$ the length of the video as the number of images and $V_i^j=(v_i,\ldots,v_{i+j-1})$ as the subsequence of $V$ starting at index $i$ having a length of $j$.

For a matrix $D$ we denote $D_{i,j}$ the entry in the $i$-th row and $j$-th column.
Consider a distance function $d:\mathbb I\times\mathbb I\rightarrow\mathbb R^{\geq 0}$.
The self-similarity matrix of a video is the matrix with all pairwise image distances:
\begin{displaymath}
    \triangle V := \left( d(v_i,v_{i+j}) \right)_{0\leq i < n-1, 1\leq j<n-i}
\end{displaymath}
i.e. $\Delta V_{i,j} = d(v_i,v_{i+j})$.
All videos are considered to have the same frame rate (i.e. the number of frames per second).

We consider a fixed dataset of videos and a query video.
CBCD is the problem to find a copy of the video in the dataset if one exists.

\subsection{Distance Function: Milestones}
\label{sec:Milestones}

The proposed CBCD decision algorithm is based on a distance function.
We like to motivate that distance function by unfolding its development step by step.

\subsubsection{Pixel Distance}
\label{MeanPixelDistance}
An exact copy of a video is identical with the original at each pixel in each frame.
Since we usually do not have exact copies in the real world, our first intuition is to measure the distance of all corresponding pixels: $\delta_1(U,V) := \sum_{0\leq i< n} d(u_i,v_i)$ where $d(u_i,v_i)$ is the sum of pixel distances of all corresponding pixels of $u_i$ and $v_i$.

\subsubsection{Delta Distance}
\label{sec:deltadistance}
The pixel distance function is not able to handle mirrored copies.
Here, our idea is inspired by the triangle inequality commonly used in metric spaces.
We use it to compare two videos indirectly, such that we are robust against this transformation.

Therefore, consider two videos $U=(u_i)_{0\leq i<n}$ and $V=(v_i)_{0\leq i<n}$ and a distance function $d$ comparing two images.
For the moment, consider $d$ to fulfill the triangle inequality, thus we obtain:
\begin{displaymath}
    d(u_i,u_j) \leq d(u_i, v_i) + d(v_i, v_j) + d(v_j, u_j)
\end{displaymath}%
\vspace{-15pt}%
\begin{displaymath}
	\Longrightarrow |d(u_i,u_j)-d(v_i,v_j)| \leq d(u_i, u_j) + d(v_i, v_j)
\end{displaymath}
Considering the two self-similarity matrices $\triangle U$ and $\triangle V$ we set $\Delta:=\Delta U-\Delta V$ and
\begin{displaymath}
    \delta_2(U,V) := \max_{1\leq j<n} \sum_{i=0}^{n-j-1} |\Delta_{i,i+j}|
\end{displaymath}
This is the crucial step which makes our approach modular.
It is easy to see, that $\delta_2(U,V) \leq 2\cdot\delta_1(U,V)$.

Remark, that the sum of pixel distances of two images does not change when mirroring both images.
Furthermore, the problem of comparing images of different resolutions does not occur further on since $\delta_2(U,V)$ only calls the distance function on images of the same video.

\subsubsection{Mean Pixel Distance}
\label{sec:meanpixeldistance}
With the last step, we have an approximation for the pixel distance function which does not depend on the image resolution.
But usually, the pixel distance function returns large values for large images and small values for small images.
Hence, it is reasonable to use a mean pixel distance function $d_2:\mathbb I\times\mathbb I\rightarrow\mathbb R^{\geq 0}$ instead.
Experiments indicated that the robustness of $d_2$ against resolution changes are inherited by the delta distance function.

\subsubsection{Different Lengths}
\label{sec:differentlengths}
Now, the distance function is robust against resolution changes.
But two videos being compared usually differ in their length.
Therefore, we use a common approach by interpreting the shorter video as a window for the longer video.

Consider $U$ and $V$ with $|U|\leq|V|$, then let
\begin{displaymath}
    \Delta_\delta(U,V) = \min_{0\leq i\leq |V|-|U|} \delta\left( U, V_i^{|U|} \right)
\end{displaymath}
where $\delta$ is some video distance function working for videos of the same length.
For videos $U$ and $V$ with $|U|>|V|$ we simply set $\Delta_\delta(U,V) = \Delta_\delta(V,U)$.

\subsubsection{Video Preprocessing}
Next, we consider blurred copies.
Since blurring is similar to decreasing and then increasing the resolution again, we simply decrease the resolution of all videos to a certain value before being processed.
Although losing information for high resolution videos, the reduced copies still have enough information to compare their content meaningfully.

The concrete resolution is a parameter for our algorithm.
It is evaluated in Section~\ref{sec:experiments}.

\subsubsection{Mean Delta Distance}
\label{sec:meandeltadistance}
Since \ref{sec:differentlengths}, our video distance function can handle videos of different lengths.
But there is a similar problem as in \ref{sec:meanpixeldistance}, i.e. short videos have small distance values, whereas long videos have large distance values.
It suffices to simply calculate a mean value over all image distance values, i.e. let $\Delta:=\Delta U-\Delta V$ and
\begin{displaymath}
    \delta_3(U,V) := \max_{1\leq j<n} \frac{1}{n-j} \sum_{i=0}^{n-j-1} |\Delta_{i,j}|
\end{displaymath}
Thus, $\Delta_{\delta_3}$ is a video distance function approximating the mean pixel distance.

\subsubsection{Black (Static) Pixels}
\label{sec:staticpixels}
Changing the aspect ratio may add black borders to videos.
On the other hand, cropping is also used to remove black borders from videos.
These black pixels have a huge impact on the mean pixel distance.

However, we already mentioned that the delta distance function inherits the robustness of the image distance function (cf.~\ref{sec:meanpixeldistance}).
Thus, it suffices to modify the image distance function such that it is robust against adding black borders.

The solution is simple: Instead of dividing by the total number of pixels, the new image distance function $d_3$ divides by the number of different pixels.
Indeed, the new distance function is more robust against removing black borders.
But it is less robust against noise, thus there is room for improvement.

\subsubsection{Brightness}
Changing the brightness is a common video transformation.
A similar effect appears when filming a video played on a screen, i.e. when making a screener.
To be more robust against brightness changes, we normalize the brightness of the videos which are to be compared to each other.
For the delta distance, this is equivalent to normalizing the rows of their self-similarity matrices:
\begin{displaymath}
	\bar\Delta U = \left( \frac{\Delta U_{i,j}}{\sum_{h=0}^{n-j-1}\Delta U_{h,j} }  \right)_{0\leq i<n-1, 1\leq j<n-i}
\end{displaymath}

\subsubsection{Performance}
\label{sec:performance}
So far, our algorithm has quadratic runtime, since the size of the self-similarity matrices is quadratic in the video length.
To reduce that complexity, we drop all but $n\cdot\log n$ entries of the matrices:
Let $\bar\Delta := \bar\Delta U - \bar\Delta V$ and
\begin{displaymath}
    \delta_4(U,V) := \max_{1\leq j<n, j\in 2^{\mathbb N}} \frac{1}{j} \sum_{i=0}^{n-j-1} |\bar\Delta_{i,j}|
\end{displaymath}
Remark, that quadratic runtime remains because of the windowing approach (cf. Section~\ref{sec:differentlengths}).

\subsection{Copy Detection Decision Algorithm}
\label{sec:decisionalgorithm}

If we put the ideas from Section~\ref{sec:Milestones} together, we achieve the \emph{MuscleDistance}:
Considering $\Delta_{\delta_4}$ with $d_3$ as image distance function, we have a video distance function which is robust against
flipping (e.g. mirroring vertically), rotating by a small angle, 
changing resolution,
adding or removing black borders, 
and blurring.

Now, the proposed decision algorithm for CBCD searches for nearest neighbors to a given query using the MuscleDistance.
The algorithm returns a positive answer, iff it found a nearest neighbor closer than a certain threshold.
The nearest neighbor and the query videos are considered copies.

\begin{enumerate}
\item 
\end{enumerate}
\section{Evaluation}
\label{experiments}
\label{sec:experiments}

With our experiments, we like to evaluate three different questions:
(1) Is our proposed distance function discriminating?
(2) Is it applicable for CBCD?
(3) What is the expected computation time?

\subsection{Discriminability} \label{sec:discriminability}
We used the TRECVID 2012 dataset \cite{TRECVID} to evaluate the discriminability.
Unfortunately, we could not use all videos, since some URLs were broken.
However, we used 7827 videos (ca. $210$ hours in total) which nearly is the complete dataset.
The videos length ranges from 11 seconds to 34 minutes.
We resampled these videos to 8 frames per second and rescaled them to a width of 132 pixels (keeping aspect ratio) using ffmpeg version N-59207-gd2d794f \cite{FFMPEG}.

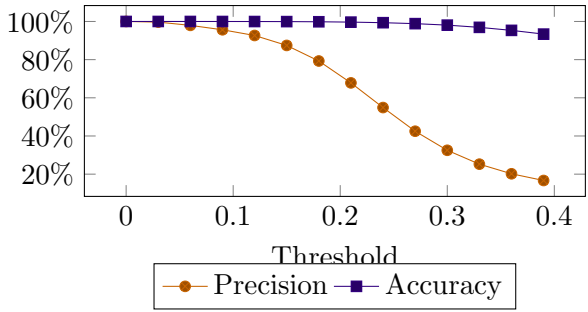
\begin{figure}[ht]
    \centering
    \begin{tikzpicture}
      \begin{axis}[
        width=.5\textwidth,
        height=0.25\textwidth,
        xlabel={Threshold},
        legend columns=3,
        yticklabel={\pgfmathparse{\tick*100}\pgfmathprintnumber{\pgfmathresult}\%},
        legend style={at={(0.5,-0.35)},anchor=north},
        cycle list name=mycolorlist,
      ]
      \addplot table[x index=0, y index=1] {images/precision.txt}; \addlegendentry{Precision};
      \addplot table[x index=0, y index=1] {images/accuracy.txt}; \addlegendentry{Accuracy};
      \end{axis}
    \end{tikzpicture}
    \caption{Mean precision and accuracy of nearest neighbor queries using our video distance.}
    \label{fig:TrecvidPrecision}
\end{figure}

For each video we ran a nearest neighbor query on the whole dataset and calculated the precision and accuracy.
Figure~\ref{fig:TrecvidPrecision} shows that the mean accuracy is above $90\%$ for all distances up to $0.4$.
Furthermore, the precision is close to $100\%$ for small thresholds, i.e. this distance function is discriminating.


\definecolor{colprecision}{RGB}{200, 100, 0}
\definecolor{colaccuracy}{RGB}{50, 0, 128}
\pgfplotscreateplotcyclelist{mycolorlist}{%
colprecision,every mark/.append style={fill=colprecision!80!black},mark=otimes*\\%
colaccuracy,every mark/.append style={fill=colaccuracy!80!black},mark=square*\\%
}

\subsection{CBCD Applicability}

We challenged the MuscleVCD ST1 benchmark using our CBCD decision algorithm (cf. Section~\ref{sec:decisionalgorithm}).
We evaluated different parameter settings for our algorithm (cf. Figure~\ref{fig:MuscleVCDScoring}).
In the preprocessing phase we took different image resolutions varying from $44$ to $220$ pixels in width and keeping the aspect ratio.
Also, we ran the benchmark with different frames per second ranging from 1 fps to 10 fps.

\pgfplotsset{
  colormap={blackwhite}{[5pt]
    rgb255(0pt)=(0, 0, 0); 
    rgb255(200pt)=(50, 0, 128); 
    rgb255(400pt)=(200, 0, 130); 
    rgb255(600pt)=(200, 100, 0); 
    rgb255(800pt)=(255, 255, 0); 
    rgb255(1000pt)=(255, 255, 255) 
},
}
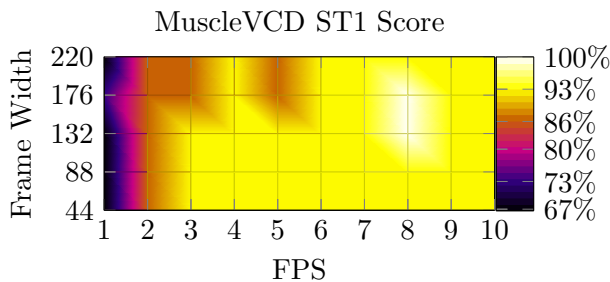
\begin{figure}[ht]
    \centering
    \begin{tikzpicture}

    \begin{axis}[
            xlabel=FPS,
            ylabel=Frame Width,
            xtick=data,
            ytick=data,
            title=MuscleVCD ST1 Score,
            width=.41\textwidth,
            height=0.22\textwidth,
            view={0}{90},
            point meta={z/15},
            colorbar sampled,
            colorbar,
            colorbar style={
            		ytick={0.67,0.73,0.8,0.86,0.93,1},
                    yticklabel={\pgfmathparse{\tick*100}\pgfmathprintnumber{\pgfmathresult}\%},
            		at={(1.1, 0.5)},
                    anchor=east,
            },
    ]
      \addplot3 [surf, shader=faceted interp]
                      file{images/muscle.txt};
    \end{axis}
    \end{tikzpicture}
    \caption{MuscleVCD Score evaluation.}
    \label{fig:MuscleVCDScoring}
\end{figure}%

\subsection{Performance Evaluation}
We present three experiments which ran on one core of an Intel(R) Xeon(R) CPU E5-2620 0 @ 2.00GHz to analyze the runtime of our approach.
Each of the experiments consists of a preprocessing (extracting the video descriptor) and a comparison phase.

The preprocessing of the TRECVID data set took about 2~hours.
Thus, we extracted about 60 video descriptors per minute.
Afterwards we compared all videos against each other as described in Section~\ref{sec:discriminability}.
Here, our approach executed 690 comparisons per second.



Next, we measured the runtime of the ST1 benchmark using 3 frames per second and $44$ pixels in width as well as $8$ frames per second and $132$ pixels in width.
The preprocessing took 40 minutes resp. 176 minutes and the comparison phase took 15 minutes resp. 108 minutes.
Table \ref{tbl:summary} summarizes the runtime and compares them to some related work also performing the MuscleVCD benchmark.

\begin{table}[ht]
    \begin{tabular}{ | r | c | c | }
        \hline
        & \textbf{Score} & \textbf{Runtime} \\
        \hline
        Yeh and Cheng \cite{fastsequencematching} & $86\%$ & $1394$ sec \\
        \hline
        Yeh and Cheng \cite{Yeh2009} & $93\%$ & ? \\
        \hline
        Poullot et al. \cite{scalablevideodbmining} w/o Index & $93\%$ & $1380$ sec \\
        \hline
        Poullot et al. \cite{scalablevideodbmining} with Index & $93\%$ & $20$ sec \\
        \hline
        Ours with $8$ fps & $100\%$ & $6508$ sec \\
        \hline
        Ours with $3$ fps & $93\%$ & $905$ sec \\
        \hline
    \end{tabular}
    \caption{Comparisons to related work}
    \label{tbl:summary}
\end{table}

\section{Conclusion and Future Work}
\label{conclusion}

We have presented a new spatio-temporal algorithm for content based video copy detection.
Our algorithm computes a descriptor inspired by the self-similarity matrix in a preprocessing phase using an image distance function on certain pairs of frames.
This image distance function can be chosen arbitrarily before preprocessing all videos.
These descriptors are then used to compare the videos.

The experiments show that our approach is robust and discriminating in most cases.
It remains future work to look for more robust respective discriminating image distance functions.
An image distance function utilizing SIFT or SURF local features is a promising approach.

In Section \ref{sec:experiments} we compared our algorithm to other approaches with respect to time complexity as well as quality.
It achieves high score in the MuscleVCD ST1 benchmark.
Furthermore, the experiments suggest that it is competitive with approaches from related work w.r.t. runtime.




\bibliographystyle{abbrv}
\bibliography{literature}

\end{document}